\documentclass[structabstract]{aa}  
\usepackage{graphicx}
\usepackage{txfonts}
\begin{document}
   \title{Colors and taxonomy of Centaurs and Trans-Neptunian Objects 
\thanks{Based on observations carried out at the European Southern 
Observatory (ESO), Chile (Programmes 178.C-0036 and 178.C-0867).}}

   \subtitle{}

   \author{D. Perna \inst{1,2,3},
	  M. A. Barucci \inst{1},
	  S. Fornasier \inst{1,4},
	  F. E. DeMeo \inst{1},
          A. Alvarez-Candal \inst{1,5},
          F. Merlin \inst{1,6},
          E. Dotto \inst{3},
          A. Doressoundiram \inst{1}, and
	  C. de Bergh \inst{1}
}

\offprints{D. Perna}

\institute{LESIA, Observatoire de Paris, 5 Place Jules Janssen,
92195 Meudon Principal Cedex, France\\
	\email{davide.perna@obspm.fr}
\and Universit\`a di Roma Tor Vergata, Italy
\and INAF-Osservatorio Astronomico di Roma, Italy
\and Universit\'{e} Paris Diderot -- Paris 7, France
\and European Southern Observatory, Chile
\and Department of Astronomy, University of Maryland, USA
}

   \date{Received ...; accepted ...}

% \abstract{}{}{}{}{} 
	% 5 {} token are mandatory
	 
\abstract
 {The study of the surface properties of Centaurs and Trans-Neptunian Objects (TNOs)
 provides essential information about the early conditions and evolution of
 the outer Solar System. Due to the faintness of most of these distant and icy
 bodies, photometry currently constitutes the best technique to survey a
 statistically significant number of them.}
 {Our aim is to investigate color properties of a large sample of minor bodies
 of the outer Solar System, and set their taxonomic classification.}
 {We carried out visible and near-infrared photometry of Centaurs and TNOs, making
 use, respectively, of the FORS2 and ISAAC instruments at the Very Large
 Telescope (European Southern Observatory).
 Using G-mode analysis, we derived taxonomic classifications according to the Barucci et al. (2005a) system.}
 {We report photometric observations of 31 objects,
 10 of them have their colors reported for the first time ever. 28 Centaurs and TNOs have been
 assigned to a taxon.}
 {We combined the entire sample of 38 objects taxonomically classified in the framework of our programme
 (28 objects from this work; 10 objects from DeMeo et al. 2009a) with previously classified TNOs and Centaurs,
 looking for correlations between taxonomy and dynamics.
 We compared our photometric results to literature data, finding hints of heterogeneity for the surfaces of 4 objects.}


\titlerunning{Colors and taxonomy of Centaurs and Trans-Neptunian Objects}
\authorrunning{D. Perna et al.}

\maketitle
	%
	%________________________________________________________________

\section{Introduction}
The investigation of the surface properties of the minor bodies in the outer
Solar System constitutes a major topic in modern planetary science, since they
represent the ``vestiges'' of the leftover planetesimals from the early
accretional phases of the outer proto-planetary disk.
Even though they are affected by space weathering and collisional evolution
(see, e.g., Hudson et al. 2008 and Leinhardt et al. 2008), they
present the most pristine material in present times available for our studies, from
which we can learn about the origin and early evolution of the Solar System at
large distances from the Sun.

Because of the faintness of Trans-Neptunian Objects (TNOs) and
Centaurs, spectroscopic observations (which could provide the most detailed
information about their surface compositions) are feasible only for a small number
of them, even when using the largest ground-based telescopes
(see Barucci et al. 2008 for a recent review).
Hence photometry is still the best tool to investigate the surface properties of
a significant sample of these objects, allowing a global view of the whole
known population.

To date photometric surveys have observed more than 200 objects.
Statistical analyses were performed to search for possible correlations
between colors and physical and orbital parameters
(see Doressoundiram et al. 2008 and Tegler et al. 2008 for recent reviews).
As the major result, a clustering of ``cold'' (low eccentricity, low inclination)
classical TNOs (see Gladman et al. 2008 for a dynamical classification of
objects in the outer Solar System) with very red colors was found.

Even if photometric colors cannot provide firm constraints on the surface composition, since
they depend also on scattering effects in particulate regoliths and viewing geometry,
they can be used to classify the objects in different groups that reasonably indicate
different composition and/or evolutional history.
A new TNO taxonomy (Barucci et al. 2005a; Fulchignoni et al. 2008) based on color indices
($B-V$, $V-R$, $V-I$, $V-J$, $V-H$, and $V-K$) identifies four classes with increasingly
red colors: BB (neutral color), BR, IR, and RR (very red).

An ESO (European Southern Observatory) Large Programme devoted to the study of TNOs
and Centaurs, by means of different techniques, was lead by M. A. Barucci in 2006-2008.
In this framework, visible and near-infrared (NIR) photometry of a total of 45 objects
was performed. The results from data acquired between October 2006 and September 2007
were published in DeMeo et al. (2009a).
In this paper we present all the photometric observations executed during the second
year of the Large Programme (November 2007 - November 2008), regarding 31 objects.
For 28 of these targets we were able to determine the Barucci et al. (2005a) taxonomic classification,
 via the G-mode statistical method presented in Fulchignoni et al. (2000),
and to compare our results with literature data whenever available.

An analysis of the entire sample (151 objects) of taxonomically classified TNOs and Centaurs
has been performed, searching for correlations between dynamical properties and taxonomy.

\begin{table*}[p]
\caption{Observational circumstances.}
\label{observations}      % is used to refer this table in the text
\centering
\begin{tabular}{l l r r c}        % centered columns (4 columns)
\hline\hline                 % inserts double horizontal lines
Object                      & Date	   & $\Delta$ (AU) & $r$ (AU) & $\alpha$ (deg)      \\
\hline                        % inserts single horizontal line
(5145) Pholus               &  12 Apr 2008 & 21.212	   & 21.864   & 2.0 \\
(10199) Chariklo            &	3 Feb 2008 & 13.311	   & 13.395   & 4.2 \\
                            &	4 Feb 2008 & 13.296	   & 13.395   & 4.2 \\
(20000) Varuna              &	4 Dec 2007 & 42.605	   & 43.391   & 0.8 \\
(42301) 2001 UR$_{163}$     &	5 Dec 2007 & 49.625	   & 50.308   & 0.8 \\
(42355) Typhon              &  12 Apr 2008 & 16.892	   & 17.650   & 2.2 \\
(44594) 1999 OX$_{3}$       &  21 Sep 2008 & 22.025	   & 22.889   & 1.3 \\
                            &  22 Sep 2008 & 22.033	   & 22.888   & 1.3 \\
(52872) Okyrhoe             &	3 Feb 2008 & 4.883	   & 5.800    & 3.9 \\
                            &	4 Feb 2008 & 4.877	   & 5.800    & 3.7 \\
(55576) Amycus              &  12 Apr 2008 & 15.205	   & 16.056   & 1.9 \\
(55637) 2002 UX$_{25}$      &	6 Dec 2007 & 41.263	   & 41.975   & 0.9 \\
(55638) 2002 VE$_{95}$      &	5 Dec 2007 & 27.297	   & 28.248   & 0.5 \\
                            &	6 Dec 2007 & 27.301	   & 28.249   & 0.6 \\
                            &  22 Nov 2008 & 27.379	   & 28.341   & 0.5 \\
                            &  23 Nov 2008 & 27.378	   & 28.341   & 0.4 \\
(73480) 2002 PN$_{34}$      &  10 Nov 2007 & 14.894	   & 15.344   & 3.3 \\
(90377) Sedna               &  21 Sep 2008 & 87.416	   & 88.015   & 0.5 \\
                            &  22 Sep 2008 & 87.402	   & 88.014   & 0.5 \\
(90482) Orcus               &	3 Feb 2008 & 46.904	   & 47.807   & 0.5 \\
                            &	4 Feb 2008 & 46.900	   & 47.807   & 0.5 \\
(120061) 2003 CO$_{1}$      &	4 Feb 2008 & 10.937	   & 11.080   & 5.1 \\
                            &  12 Apr 2008 & 10.179	   & 11.123   & 1.8 \\
                            &  13 Apr 2008 & 10.175	   & 11.123   & 1.8 \\
(120132) 2003 FY$_{128}$    &  12 Apr 2008 & 37.477	   & 38.454   & 0.3 \\
(120178) 2003 OP$_{32}$     &  21 Sep 2008 & 40.546	   & 41.365   & 0.8 \\
(120348) 2004 TY$_{364}$    &  22 Nov 2008 & 38.840	   & 39.591   & 0.9 \\
                            &  23 Nov 2008 & 38.849	   & 39.591   & 1.0 \\
(136199) Eris               &	7 Dec 2007 & 49.652	   & 50.310   & 0.8 \\
(144897) 2004 UX$_{10}$     &	4 Dec 2007 & 38.145	   & 38.837   & 1.0 \\
                            &	5 Dec 2007 & 38.158	   & 38.837   & 1.1 \\
                            &	6 Dec 2007 & 38.170	   & 38.837   & 1.1 \\
                            &  23 Nov 2008 & 38.059	   & 38.879   & 0.8 \\
(145451) 2005 RM$_{43}$     &	4 Dec 2007 & 34.298	   & 35.195   & 0.7 \\
                            &	7 Dec 2007 & 34.316	   & 35.196   & 0.7 \\
(145453) 2005 RR$_{43}$     &	4 Dec 2007 & 37.623	   & 38.511   & 0.6 \\
                            &	7 Dec 2007 & 37.642	   & 38.512   & 0.7 \\
(174567) 2003 MW$_{12}$     &  12 Apr 2008 & 47.314	   & 47.968   & 0.9 \\
                            &  13 Apr 2008 & 47.302	   & 47.968   & 0.9 \\
(208996) 2003 AZ$_{84}$     &  22 Nov 2008 & 44.879	   & 45.458   & 1.0 \\
                            &  23 Nov 2008 & 44.865	   & 45.457   & 1.0 \\
2002 KY$_{14}$              &  21 Sep 2008 & 7.802	   & 8.649    & 3.8 \\
                            &  22 Sep 2008 & 7.810	   & 8.649    & 3.8 \\
2003 UZ$_{117}$             &  22 Nov 2008 & 38.420	   & 39.368   & 0.4 \\
                            &  23 Nov 2008 & 38.423	   & 39.367   & 0.4 \\
2003 UZ$_{413}$             &	4 Dec 2007 & 41.171	   & 42.004   & 0.7 \\
                            &  21 Sep 2008 & 41.466	   & 42.163   & 1.0 \\
                            &  22 Nov 2008 & 41.276	   & 42.197   & 0.5 \\
                            &  23 Nov 2008 & 41.282	   & 42.198   & 0.5 \\
2007 UK$_{126}$             &  21 Sep 2008 & 45.131	   & 45.618   & 1.1 \\
                            &  22 Sep 2008 & 45.117	   & 45.617   & 1.1 \\
2007 UM$_{126}$             &  21 Sep 2008 & 10.202	   & 11.163   & 1.5 \\
                            &  22 Sep 2008 & 10.199	   & 11.165   & 1.5 \\
2007 VH$_{305}$             &  22 Nov 2008 & 7.854	   & 8.638    & 4.2 \\
                            &  23 Nov 2008 & 7.863	   & 8.636    & 4.3 \\
2008 FC$_{76}$              &  20 Sep 2008 & 10.968	   & 11.690   & 3.5 \\
                            &  21 Sep 2008 & 10.976	   & 11.688   & 3.6 \\
2008 SJ$_{236}$             &  22 Nov 2008 & 5.522	   & 6.364    & 5.0 \\
                            &  23 Nov 2008 & 5.530	   & 6.363    & 5.1 \\ 
\hline                                   %inserts single line
\end{tabular}
~\\
\raggedright
\smallskip
NOTE: $\Delta$, $r$ and $\alpha$ are the topocentric distance, the heliocentric distance, and the phase angle, respectively.\\
\end{table*}

\section{Observations and data reduction}
All of the data presented in this work were obtained with the
ESO Unit Telescope 1 (Antu) of the Very Large Telescope (VLT),
located in Cerro Paranal, Chile.

The observational circumstances are reported in Table~\ref{observations}.

\subsection{Visible}
Visible photometry was performed with the FORS2 instrument (Appenzeller et al. 1998), equipped with a
mosaic of two 2000$\times$4000 MIT CCD with square 15 $\mu$m pixels.
The observations were carried out with the standard resolution (SR) collimator
and a 2$\times$2 binning, yielding a resolution of 0.25 arcsec/pixel.
We used the broadband V, R, I filters, centered at 0.554, 0.655 and 0.768 $\mu$m,
respectively, adjusting the exposure time according to the object magnitude.

The images were reduced using standard procedures with the
MIDAS software package: after subtraction of the bias from the raw data and
flat-field correction, the instrumental magnitudes were measured
via aperture photometry, with an integrating radius typically about three times
the average seeing and sky subtraction performed using a 5-10 pixel wide
annulus around each object.
The aperture correction method (see, e.g., Barucci et al. 2000)
was used for only a few cases (faint target and/or nearby field stars) to determine the object flux.
The absolute calibration of the magnitudes was obtained by means of the
observation of several Landolt (1992) standard fields.

\subsection{Near-infrared}
NIR photometry was performed with the ISAAC instrument (Moorwood et al. 1998), in
SWI1 (short wavelength imaging) mode, using the 1024$\times$1024 Hawaii Rockwell
array with a pixel size of 18.5 $\mu$m and a scale of 0.148 arcsec/pixel.
We observed with the J, H, Ks filters, with central wavelength of 1.25, 1.65
and 2.16 $\mu$m, respectively.
As is typical for NIR observations, the expositions were split in several
images with short exposure times, in order to minimize the sky background noise.

The data were pre-reduced (dark subtraction, flat-field correction, bad pixel cleaning,
sky subtraction, recombination of the images) by using the ESO ISAAC pipeline.
Then, as for the visible images, target fluxes were mostly measured (with MIDAS) using
classical photometry methods with apertures determined by the seeing and
growth curves of the objects, reserving the aperture correction method to a few cases.
To calibrate the instrumental magnitudes, standard stars from different catalogues
(Persson et al. 1998; Hawarden et al. 2001) were observed.

\section{Results}
During the second year of our ESO Large Programme
we obtained visible and NIR photometric measurements for 31 objects,
10 of them have their colors reported for the first time ever. Table~\ref{magnitudes}
lists the resulting magnitude values, as well as the computed absolute magnitudes $H$
of the targets, calculated as
\begin{equation}
H = V - 5 \log \left( r \Delta \right) - \alpha \beta,
\end{equation}
where $V$ represents the visible magnitude reported in column 4 of
Table~\ref{magnitudes}, $\Delta$, $r$ and $\alpha$ are the topocentric and heliocentric
distances and the phase angle given in Table~\ref{observations}, respectively,
and $\beta$ is the phase curve slope (mag/deg). For TNOs, we assumed $\beta = 0.14\pm0.03$ mag/deg,
the modal value of the measurements published by Sheppard \& Jewitt (2002). For Centaurs and Jupiter-coupled objects,
we assumed $\beta = 0.11\pm0.01$ mag/deg, the result of a least squares fit by Doressoundiram et al. (2005)
of the linear phase function $\phi$($\alpha$) = 10$^ {- \alpha\beta}$ of data from Bauer et al. (2003).

   \begin{figure}
   \centering
   \includegraphics[angle=0,width=9cm]{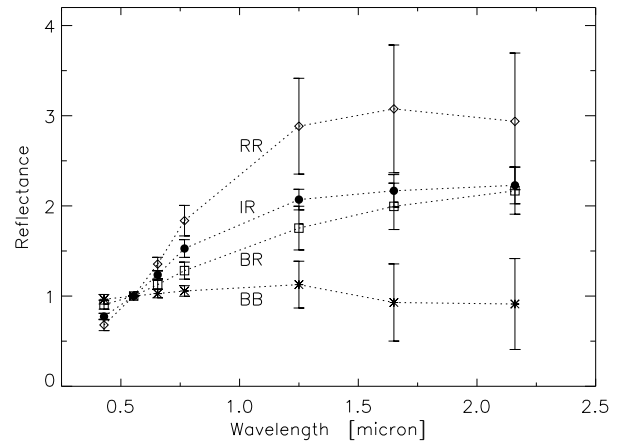}
      \caption{Average reflectance values for each taxon, normalized to the Sun and to the V colors.}
         \label{taxa}
   \end{figure}

   \begin{figure}
   \centering
   \includegraphics[angle=0,width=9cm]{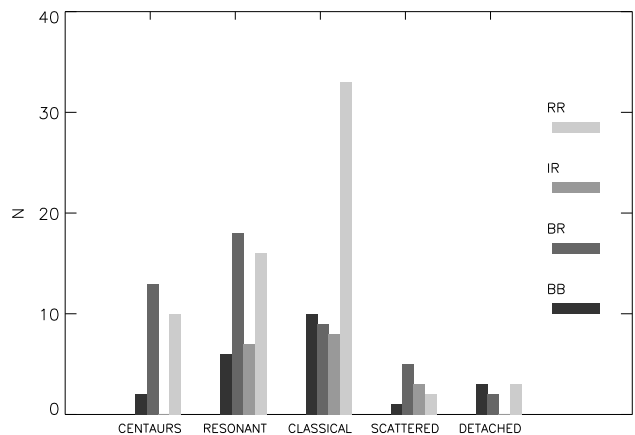}
      \caption{Distribution of the taxa within each dynamical class, as defined by Gladman et al. (2008).}
         \label{classi}
   \end{figure}

   \begin{figure}
   \centering
   \includegraphics[angle=0,width=9cm]{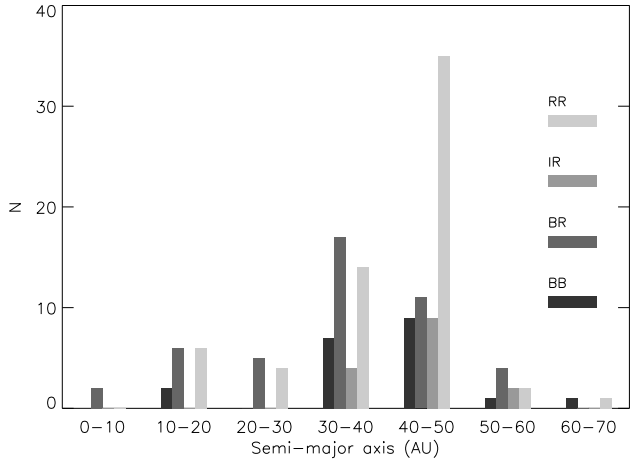}
      \caption{Distribution of the taxa with respect to the semimajor axis of the objects. A 10 AU binning is adopted.}
         \label{semiasse}
   \end{figure}

   \begin{figure}
   \centering
   \includegraphics[angle=0,width=9cm]{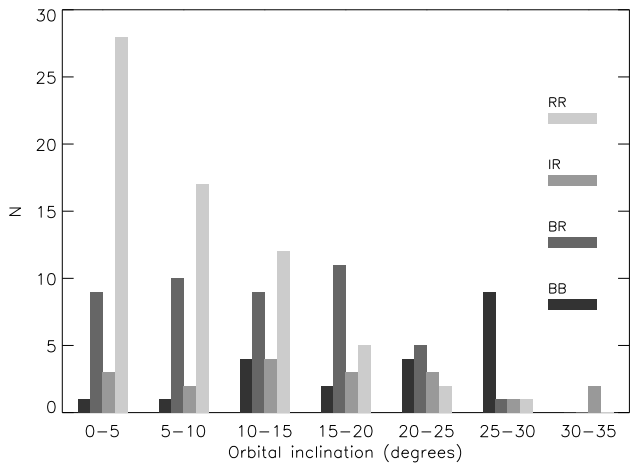}
      \caption{Distribution of the taxa with respect to the orbital inclination of the objects. A 5$\degr$ binning is adopted.}
         \label{inclinazione}
   \end{figure}

On the basis of the obtained color indices, the taxonomic classification of the targets was derived
via the G-mode statistical method presented in Fulchignoni et al. (2000),
using the taxonomy for TNOs and Centaurs introduced by Barucci et al. (2005a).
This taxonomy identifies four classes, that reasonably indicate different composition and/or evolutional history,
with increasingly red colors: BB (neutral colors with respect to the Sun), BR, IR, RR (very red colors).
We applied the above-mentioned algorithm to objects for which two or more color indices were available
(i.e., to 29 out of the 31 observed TNOs and Centaurs).
We classified each object whenever its colors were within 3$\sigma$ of the class' average values.
Obviously, a higher number of available colors implies a better reliability of the class determination.
In cases where more than one class is within 3$\sigma$, we assigned a multiple designation to the object, with taxonomic classes ordered by
ascending deviation of its colors from the class' averages.
The taxonomic designations
are reported in Table~\ref{taxonomy}, along with the dynamical classification of the objects
(according to Gladman et al. 2008).

\begin{table*}[p]
\caption{Observed magnitudes.}
\label{magnitudes}      % is used to refer this table in the text
\centering
\scriptsize{
\begin{tabular}{l l c c c c c c c c}        % centered columns (4 columns)
\hline\hline                 % inserts double horizontal lines
Object                           & Date        & UT$_{START}$  & V              & R              & I              & J              & H              & K$_{s}$        & H$_{V}$        \\
\hline                        % inserts single horizontal line
(5145) Pholus                    & 12 Apr 2008 & 05:48         & $21.33\pm0.09$ & $20.64\pm0.09$ & $19.95\pm0.15$ & ...            & ...            & ...            & $7.78\pm0.09$  \\
(10199) Chariklo$^a$             &  3 Feb 2008 & 07:41         & ...            & ...            & ...            & $17.11\pm0.06$ & $16.61\pm0.05$ & $16.38\pm0.06$ & ...            \\
                                 &  4 Feb 2008 & 07:41         & $18.79\pm0.02$ & $18.34\pm0.02$ & $17.88\pm0.03$ & ...            & ...            & ...            & $7.07\pm0.04$  \\
(20000) Varuna                   &  4 Dec 2007 & 06:09         & $20.49\pm0.03$ & $19.88\pm0.03$ & ...            &  ...           &  ...           & ...            & $4.04\pm0.04$  \\
                                 &  4 Dec 2007 & 06:12         & $20.46\pm0.04$ & $19.88\pm0.03$ & ...            & ...            & ...            & ...            & $4.01\pm0.05$  \\
(42301) 2001 UR$_{163}$          &  5 Dec 2007 & 00:52         & $21.82\pm0.06$ & $20.98\pm0.07$ & ...            & ...            & ...            & ...            & $4.72\pm0.06$  \\
                                 &  5 Dec 2007 & 00:55         & $21.81\pm0.05$ & $20.98\pm0.07$ & ...            & ...            & ...            & ...            & $4.71\pm0.06$  \\
(42355) Typhon$^b$               & 12 Apr 2008 & 00:59         & $20.50\pm0.08$ & $20.00\pm0.08$ & $19.70\pm0.08$ & ...            & ...            & ...            & $7.77\pm0.11$  \\
(44594) 1999 OX$_{3}$            & 21 Sep 2008 & 04:07         & $21.26\pm0.05$ & $20.54\pm0.06$ & $19.96\pm0.07$ & ...            & ...            & ...            & $7.57\pm0.06$  \\
                                 & 22 Sep 2008 & 04:49         & ...            & ...            & ...            & $19.08\pm0.12$ & $18.75\pm0.08$ & $18.69\pm0.08$ & ...            \\
(52872) Okyrhoe$^c$              &  3 Feb 2008 & 06:41         & ...            & ...            & ...            & $16.84\pm0.06$ & $16.39\pm0.07$ & $16.28\pm0.05$ & ...            \\
                                 &  4 Feb 2008 & 06:18         & $18.63\pm0.02$ & $18.14\pm0.02$ & $17.64\pm0.02$ & ...            &  ...           & ...            & $10.97\pm0.04$ \\
(55576) Amycus                   & 12 Apr 2008 & 03:33         & $20.42\pm0.08$ & $19.78\pm0.08$ & $19.20\pm0.15$ & ...            &  ...           & ...            & $8.27\pm0.08$  \\                    
(55637) 2002 UX$_{25}$           &  6 Dec 2007 & 01:19         & ...            & ...            &  ...           & $18.55\pm0.03$ & $18.25\pm0.04$ & $18.21\pm0.06$ & ...            \\
(55638) 2002 VE$_{95}$           &  5 Dec 2007 & 03:35         & $20.31\pm0.03$ & $19.59\pm0.04$ & ...            & ...            & ...            & ...            & $5.80\pm0.03$  \\
                                 &  5 Dec 2007 & 03:38         & $20.31\pm0.03$ & $19.59\pm0.04$ & ...            & ...            & ...            & ...            & $5.80\pm0.03$  \\
                                 &  6 Dec 2007 & 02:09         & ...            & ...            & ...            & $18.11\pm0.04$ & $17.78\pm0.04$ & $17.74\pm0.04$ & ...            \\
                                 & 22 Nov 2008 & 05:36         & $20.28\pm0.06$ & $19.53\pm0.08$ & $18.77\pm0.09$ & ...            & ...            & ...            & $5.76\pm0.06$  \\
                                 & 23 Nov 2008 & 05:57         & ...            & ...            & ...            & $18.04\pm0.07$ & $17.68\pm0.07$ & $17.63\pm0.09$ & ...            \\
(73480) 2002 PN$_{34}$$^c$       & 10 Nov 2007 & 00:36         & $20.68\pm0.03$ & $20.25\pm0.05$ & ...            & $19.00\pm0.05$ & $18.49\pm0.06$ & $18.29\pm0.05$ & $8.42\pm0.10$  \\
(90377) Sedna                    & 21 Sep 2008 & 06:33         & $21.34\pm0.04$ & $20.57\pm0.05$ & $19.93\pm0.05$ & ...            & ...            & ...            & $1.84\pm0.04$  \\
                                 & 22 Sep 2008 & 06:57         & ...            & ...            & ...            & $19.20\pm0.04$ & $18.78\pm0.06$ & ...            & ...            \\
(90482) Orcus$^c$                &  3 Feb 2008 & 04:46         & ...            & ...            & ...            & $17.91\pm0.05$ & $17.72\pm0.07$ & $17.89\pm0.05$ & ...            \\
                                 &  4 Feb 2008 & 05:04         & $19.12\pm0.02$ & $18.73\pm0.02$ & $18.37\pm0.02$ & ...            & ...            & ...            & $2.30\pm0.03$  \\
(120061) 2003 CO$_{1}$           &  4 Feb 2008 & 08:17         & $19.93\pm0.02$ & $19.45\pm0.03$ & $19.01\pm0.03$ & ...            & ...            & ...            & $8.95\pm0.05$  \\
                                 & 12 Apr 2008 & 04:40         & $19.63\pm0.08$ & $19.20\pm0.09$ & $18.82\pm0.15$ & ...            &...             & ...            & $9.16\pm0.08$  \\
(120132) 2003 FY$_{128}$         & 12 Apr 2008 & 02:22         & $20.93\pm0.09$ & $20.34\pm0.09$ & $19.86\pm0.15$ & ...            & ...            & ...            & $5.09\pm0.09$  \\
{\bf(120178) 2003 OP$_{32}$}     & 21 Sep 2008 & 02:10         & $20.25\pm0.03$ & $19.86\pm0.05$ & $19.50\pm0.05$ & ...            & ...            & ...            & $4.02\pm0.04$  \\
(120348) 2004 TY$_{364}$         & 22 Nov 2008 & 04:32         & $20.64\pm0.03$ & $20.04\pm0.04$ & $19.52\pm0.04$ & ...            & ...            & ...            & $4.58\pm0.04$  \\
                                 & 23 Nov 2008 & 03:22         & ...            & ...            & ...            & $18.87\pm0.03$ & $18.42\pm0.05$ & $18.39\pm0.08$ & ...            \\
(136199) Eris$^d$                &  7 Dec 2007 & 00:20         & ...            & ...            & ...            & $17.90\pm0.06$ & $17.85\pm0.05$ & $18.15\pm0.06$ & ...            \\  
{\bf(144897) 2004 UX$_{10}$}     &  4 Dec 2007 & 00:53         & $20.61\pm0.04$ & $20.04\pm0.04$ & ...            & ...            & ...            & ...            & $4.62\pm0.05$  \\
                                 &  4 Dec 2007 & 00:56         & $20.63\pm0.04$ & $20.05\pm0.04$ & ...            & ...            & ...            & ...            & $4.64\pm0.05$  \\
                                 &  5 Dec 2007 & 02:24         & $20.63\pm0.03$ & $20.06\pm0.04$ & ...            & ...            & ...            & ...            & $4.62\pm0.04$  \\
                                 &  6 Dec 2007 & 00:39         & ...            & ...            & ...            & $18.97\pm0.06$ & $18.55\pm0.08$ & $18.55\pm0.09$ & ...            \\
                                 & 23 Nov 2008 & 02:26         & ...            & ...            & ...            & $19.02\pm0.03$ & $18.60\pm0.04$ & $18.64\pm0.06$ & ...            \\
(145451) 2005 RM$_{43}$          &  4 Dec 2007 & 03:22         & $20.04\pm0.04$ & $19.66\pm0.03$ & ...            & ...            & ...            & ...            & $4.53\pm0.05$  \\
                                 &  4 Dec 2007 & 03:25         & $20.07\pm0.04$ & $19.66\pm0.03$ & ...            & ...            &  ...           & ...            & $4.56\pm0.05$  \\
                                 &  7 Dec 2007 & 03:14         & ...            & ...            & ...            & $18.95\pm0.04$ & $18.76\pm0.05$ & $18.71\pm0.06$ & ...            \\
(145453) 2005 RR$_{43}$          &  4 Dec 2007 & 02:26         & $20.05\pm0.03$ & $19.66\pm0.04$ & ...            & ...            & ...            & ...            & $4.16\pm0.03$  \\
                                 &  4 Dec 2007 & 02:29         & $20.08\pm0.03$ & $19.66\pm0.04$ & ...            & ...            & ...            & ...            & $4.19\pm0.03$  \\
                                 &  7 Dec 2007 & 02:24         & ...            & ...            & ...            & $19.28\pm0.04$ & $19.47\pm0.07$ & $19.67\pm0.15$ & ...            \\
{\bf(174567) 2003 MW$_{12}$}     & 12 Apr 2008 & 07:07         & $20.57\pm0.08$ & $19.99\pm0.08$ & $19.57\pm0.15$ & ...            & ...            & ...            & $3.66\pm0.08$  \\
(208996) 2003 AZ$_{84}$          & 22 Nov 2008 & 06:07         & $20.46\pm0.03$ & $20.08\pm0.04$ & $19.71\pm0.03$ & ...            & ...            & ...            & $3.77\pm0.04$  \\
                                 & 23 Nov 2008 & 07:10         & ...            & ...            & ...            & $19.26\pm0.04$ & $18.92\pm0.07$ & ...            & ...            \\
{\bf2002 KY$_{14}$}              & 21 Sep 2008 & 00:56         & $19.93\pm0.02$ & $19.23\pm0.03$ & $18.58\pm0.04$ & ...            & ...            & ...            & $10.37\pm0.04$ \\
                                 & 22 Sep 2008 & 02:27         & ...            & ...            & ...            & $17.85\pm0.06$ & $17.56\pm0.07$ & $17.54\pm0.08$ & ...            \\ 
2003 UZ$_{117}$                  & 22 Nov 2008 & 03:21         & $21.13\pm0.03$ & $20.77\pm0.04$ & $20.43\pm0.04$ & ...            & ...            & ...            & $5.18\pm0.03$  \\
                                 & 23 Nov 2008 & 05:08         & ...            & ...            & ...            & $20.41\pm0.09$ & $20.63\pm0.17$ & ...            & ...            \\
{\bf2003 UZ$_{413}$}             &  4 Dec 2007 & 04:28         & $20.70\pm0.04$ & $20.22\pm0.04$ & ...            & ...            & ...            & ...            & $4.41\pm0.05$  \\
                                 &  4 Dec 2007 & 04:31         & $20.67\pm0.04$ & $20.22\pm0.04$ & ...            & ...            & ...            & ...            & $4.38\pm0.05$  \\
                                 & 21 Sep 2008 & 05:31         & $20.71\pm0.04$ & $20.25\pm0.05$ & $19.88\pm0.06$ & ...            & ...            & ...            & $4.36\pm0.05$  \\
                                 & 22 Nov 2008 & 03:33         & $20.63\pm0.03$ & $20.22\pm0.04$ & $19.82\pm0.04$ & ...            & ...            & ...            & $4.36\pm0.03$  \\
                                 & 23 Nov 2008 & 04:03         & ...            & ...            & ...            & $19.29\pm0.05$ & $18.92\pm0.09$ & $18.77\pm0.09$ & ...            \\
{\bf2007 UK$_{126}$}             & 21 Sep 2008 & 08:05         & $20.41\pm0.03$ & $19.79\pm0.04$ & $19.32\pm0.04$ & ...            & ...            & ...            & $3.69\pm0.04$  \\
                                 & 22 Sep 2008 & 07:43         & ...            &  ...           & ...            & $18.88\pm0.07$ & $18.52\pm0.08$ & ...            & ...            \\ 
{\bf2007 UM$_{126}$}             & 21 Sep 2008 & 04:26         & $20.88\pm0.03$ & $20.44\pm0.03$ & $20.00\pm0.03$ & ...            & ...            & ...            & $10.43\pm0.03$ \\
                                 & 22 Sep 2008 & 05:57         & ...            & ...            & ...            & ...            & $18.84\pm0.13$ & $18.50\pm0.08$ & ...            \\ 
{\bf2007 VH$_{305}$}             & 22 Nov 2008 & 02:14         & $21.44\pm0.04$ & $20.96\pm0.04$ & $20.48\pm0.05$ & ...            & ...            & ...            & $11.82\pm0.06$ \\
                                 & 23 Nov 2008 & 00:38         & ...            & ...            & ...            & $19.70\pm0.10$ & $19.06\pm0.12$ & $19.43\pm0.15$ & ...            \\  
{\bf2008 FC$_{76}$}              & 20 Sep 2008 & 23:55         & $20.38\pm0.03$ & $19.67\pm0.05$ & $19.04\pm0.06$ & ...            & ...            & ...            & $9.46\pm0.05$  \\
                                 & 21 Sep 2008 & 23:53         & ...            & ...            &  ...           & $18.40\pm0.08$ & $18.00\pm0.08$ & $17.88\pm0.09$ & ...            \\
{\bf2008 SJ$_{236}$}             & 22 Nov 2008 & 00:24         & $20.75\pm0.03$ & $20.13\pm0.03$ & $19.63\pm0.04$ & ...            & ...            & ...            & $12.47\pm0.06$ \\
                                 & 23 Nov 2008 & 01:28         & ...            & ...            & ...            & $19.01\pm0.07$ & $18.60\pm0.11$ & $18.83\pm0.12$ & ...            \\
\hline                                   %inserts single line
\end{tabular}}
~\\
\raggedright
\smallskip
\small
NOTE: Objects in bold have their colors reported for the first time ever.\\
$^a$ Computed magnitudes from Guilbert et al. (2009).\\
$^b$ Computed magnitudes from Alvarez-Candal et al. (2009).\\
$^c$ Computed magnitudes from DeMeo et al. (2009b).\\
$^d$ Computed magnitudes from Merlin et al. (2009).\\
\end{table*}

\begin{table*}[p]
\caption{Taxonomic classification.}
\label{taxonomy}      % is used to refer this table in the text
{\centering
\begin{tabular}{l l | c c c | l}        % centered columns (4 columns)
\hline\hline                 % inserts double horizontal lines
Object                       & Dyn. Class               &                           & Taxonomy             &                  & N         \\
                             &                          & Fulchignoni et al. (2008) & DeMeo et al. (2009a) & This work        &           \\
\hline                        % inserts single horizontal line
(5145) Pholus                & Centaur        		& RR 		       	    & ...		   & RR		      & 2	  \\
(10199) Chariklo             & Centaur        	 	& BR 		       	    & BR,BB  	           & BR		      & 5	  \\
(42355) Typhon               & Scattered      	 	& BR 		      	    & BR		   & BR,BB	      & 5$^a$     \\
(44594) 1999 OX$_{3}$        & Scattered      	 	& RR 		      	    & ...		   & RR		      & 5	  \\
(52872) Okyrhoe              & Jupiter-coupled	 	& BR 		            & ...		   & BR,IR	      & 5	  \\
(55576) Amycus               & Centaur        	 	& RR 		            & ...		   & RR,IR	      & 2	  \\
(55637) 2002 UX$_{25}$       & Classical      	 	& IR 		            & RR,IR  	           & RR		      & 5$^b$     \\
(55638) 2002 VE$_{95}$       & Resonant (3:2) 	 	& RR 		            & ...		   & RR		      & 5	  \\
(73480) 2002 PN$_{34}$       & Scattered      	 	& ...		            & ...		   & BR,BB	      & 4	  \\
(90377) Sedna                & Detached       	 	& RR 		            & ...		   & RR		      & 4	  \\
(90482) Orcus                & Resonant (3:2) 	 	& BB 		            & ...		   & BB		      & 5	  \\
(120061) 2003 CO$_{1}$       & Centaur        	 	& ...		            & ...		   & BR		      & 2	  \\
(120132) 2003 FY$_{128}$     & Detached       	 	& ...		            & BR		   & BR		      & 5$^a$     \\
(120178) 2003 OP$_{32}$      & Classical      	 	& ...		            & ...		   & BB,BR	      & 2	  \\
(120348) 2004 TY$_{364}$     & Classical      	 	& ...		            & ...		   & IR,RR,BR	      & 5	  \\
(136199) Eris                & Detached       	 	& BB 		            & BB		   & BB		      & 5$^b$     \\
(144897) 2004 UX$_{10}$      & Classical      	 	& ...		            & ...		   & BR		      & 4	  \\
(145451) 2005 RM$_{43}$      & Detached       	 	& ...		            & BB		   & BB		      & 4	  \\
(145453) 2005 RR$_{43}$      & Classical      	 	& ...		            & BB		   & BB		      & 4	  \\
(174567) 2003 MW$_{12}$      & Classical      	 	& ...		            & ...		   & IR,BR,RR	      & 2	  \\
(208996) 2003 AZ$_{84}$      & Resonant (3:2) 	 	& BB 		            & BB		   & BB		      & 4	  \\
2002 KY$_{14}$               & Centaur        	 	& ...		            & ...		   & RR		      & 5	  \\
2003 UZ$_{117}$              & Classical      	 	& ...		            & ...		   & BB		      & 4	  \\
2003 UZ$_{413}$              & Resonant (3:2) 	 	& ...		            & ...		   & BB		      & 5	  \\
2007 UK$_{126}$              & Scattered      	 	& ...		            & ...		   & *		      & 4	  \\
2007 UM$_{126}$              & Centaur        	 	& ...		            & ...		   & BR,BB	      & 4	  \\
2007 VH$_{305}$              & Centaur        	 	& ...		            & ...		   & BR		      & 5	  \\
2008 FC$_{76}$               & Centaur        	 	& ...		            & ...		   & RR		      & 5	  \\
2008 SJ$_{236}$              & Centaur        	 	& ...		            & ...		   & RR		      & 5	  \\
\hline
(28978) Ixion                & Resonant (3:2)		& IR  		            & BB		   &  		      &	          \\
(32532) Thereus              & Centaur			& BR  		            & BB		   &  		      &	          \\
(47171) 1999 TC$_{36}$       & Resonant (3:2)		& RR  		            & RR		   &    	      &	          \\
(47932) 2000 GN$_{171}$      & Resonant (3:2)		& IR  		            & BR,IR		   &  		      &	          \\
(50000) Quaoar               & Classical		& ...		            & RR		   &  	 	      &	          \\
(54598) Bienor               & Centaur		  	& BR  		            & BR		   &  		      &	          \\
(55565) 2002 AW$_{197}$      & Classical		& IR  		            & IR,RR		   &  		      &	          \\
(60558) Echeclus             & Jupiter-coupled		& BR  		            & BR,BB		   &  		      &	          \\
(90568) 2004 GV$_{9}$        & Classical		& ...		            & BR		   &  		      &	          \\
2003 QW$_{90}$               & Classical		& ...		            & IR,RR,BR	           &  		      &	          \\
\hline                                   %inserts single line
\end{tabular}
~\\
\smallskip
%\raggedright
}  % to avoid centering the following
NOTE: Objects from the ESO Large Programme taken into account for the statistical analysis of taxa.
Dynamical classes are according to Gladman et al. (2008).
First 29 bodies are classified in this work, last 10 objects come from DeMeo et al. (2009a).
Whenever multiple taxonomic types are possible,
classes are listed by ascending deviation of the object colors from the class' averages.
The symbol * indicates that the object did not fall within any of the four taxonomic classes.
N is the number of colors we used in classifying each object.\\
$^a$ Using near-infrared data from DeMeo et al. (2009a).\\
$^b$ Using visible data from DeMeo et al. (2009a).\\
\end{table*}

We classified 28 objects: seven of them turned out to belong to the BB class,
five were BR, eight were RR. The remaining eight targets got a multiple designation.
Whenever a previous classification was available in the literature, a consistent result was obtained,
even for objects with only 2 colors.

Only 2007 UK$_{126}$, even with four color measurements,
did not fall within any class of the existing taxonomy.
Indeed, according to its visible colors an IR, RR classification could be derived,
while its infrared colors match those of a BB, BR object.
Interestingly, the same result was obtained by DeMeo et al. (2009a) for two other TNOs, (26375) 1999 DE$_{9}$ and 145452 2005 RN$_{43}$.
This fact, as well as the presence of several multiple classifications, could support the idea that
further groups could be found as the number of analysed objects increases,
leading to a refinement of the current taxonomy.

Below, we discuss selected objects in further detail.

~

\noindent {\it (10199) Chariklo}:
Using five color indices, we classified this Centaur as a BR object, as did Fulchignoni et al. (2008)
on the basis of the mean colors published in literature. Our results, 
obtained in February 2008 (and already published in Guilbert et al. 2009), agree with these average measurements
except for the $V-K$ color which is about 0.2 magnitudes redder in our dataset.
Interestingly, Chariklo was already observed in the framework of our programme in March 2007
(DeMeo et al. 2009a, who assigned to Chariklo a BR-BB classification),
but we find no match with these previous results. Since the acquisition and the reduction
of the data were carried out in the same way in both observing runs, this is a probable
indication of heterogeneity on the surface of this Centaur, as already proposed in previous works
(see Guilbert et al. 2009, and references therein).

~

\noindent {\it (90377) Sedna}:
The photometric colors and taxonomic classification (RR) we derived for Sedna are in agreement with 
those published by Barucci et al. (2005b), except for
the $V-J$ color which is approximately 0.2 magnitudes bluer in our case.
Since images in different filters have not been acquired simultaneously, the observation of different rotational
phases of the object could affect color determinations. Nevertheless the
light curve of Sedna has an amplitude of only 0.02 magnitudes (Gaudi et al. 2005),
hence different observed silhouettes of the body cannot explain the found discrepancy,
which could instead be attributed to surface heterogeneity.

~

\noindent {\it (120348) 2004 TY$_{364}$}:
Even if a triple designation (IR, RR, BR) has been assigned to this object, we note that
the IR classification is strongly favored.

~

\noindent {\it (145451) 2005 RM$_{43}$}:
The new data presented here confirm the BB classification already obtained in the framework
of our Large Programme,
even if we found both $V-J$ and $V-H$ colors $\sim0.2$ magnitudes redder than in DeMeo et al. (2009a),
while the $V-K$ color is consistent.
Observations with different filters have not been carried out simultaneously, but, as for Sedna,
the amplitude of the light curve is smaller than the observed discrepancies ($\Delta m=0.12\pm0.05$ magnitudes; Perna et al. 2009),
suggesting possible surface heterogeneity. 

~

\noindent {\it (145453) 2005 RR$_{43}$}:
We confirm that 2005 RR$_{43}$ is a BB object, as classified by DeMeo et al. (2009a), but
our $V-H$ color is 0.4 magnitudes bluer than published by the same authors.
Again, images with different filters have not been acquired at the same moment but
the amplitude of the light curve is only $\Delta m=0.12\pm0.03$ magnitudes (Perna et al. 2009),
so the observation of different compositions on the surface is a likely explanation for the
reported discrepancy.

~

\noindent {\it (174567) 2003 MW$_{12}$}:
Although this object was classified as IR, BR, RR by our analysis, as for 2004 TY$_{364}$ the IR designation is strongly favored.

\section{Statistical analysis}
In the framework of our ESO Large Programme we derived the taxonomy of 38 objects
(28 objects from this work plus 10 objects from DeMeo et al. 2009a; see Table~\ref{taxonomy}).
Nineteen of them have been classified for the first time, while 4 targets were assigned different
classes with respect to the results by Fulchignoni et al. (2008),
who classified all of the 133 TNOs and Centaurs for which data from the literature were available before our observations.

Considering a total sample of 151 objects
(because 1998 WU$_{24}$ has an unusual orbit, it is not a Centaur, and therefore is not considered in this analysis),
we obtained the average colors of each taxon, reported in Table~\ref{colors} and
represented in Fig.~\ref{taxa} as reflectance values normalized to the Sun and to the V colors.
In the cases where multiple taxonomic classes were assigned to an object, we took into account only the first designation.
Then, we analysed the distribution of the four taxonomic groups with respect to the dynamical properties of the objects.

First of all, we verified the sampling of the taxa within each dynamical class (Fig.~\ref{classi}). The
well-known color bimodality of Centaurs (see, e.g., Peixinho et al. 2003) clearly emerges, since 13 out of 25 objects belong
to the BR group, while 10 of them fall in the RR class.
All of the four new IR-classified objects are classical TNOs, confirming the finding that IR objects
seem to be concentrated in the resonant and classical dynamical classes,
as stated by Fulchignoni et al. (2008).
As reported by the same authors, RR objects dominate the classical TNOs. Our new results, however, do not
conform to this behavior, as a quite equal division of the four taxa appears among the classical TNOs in the objects
constituting the Large Programme sample.

In Fig.~\ref{semiasse} we present the distribution of taxonomical classes with respect to the semimajor axes $a$ of the objects.
A 10 AU binning is adopted (nine objects are out of the scale).
As already noted by Fulchignoni et al. (2008), the more distant TNOs belong to all the four taxa in a quite uniform way,
while for $a \la 30\degr$ the BR and RR classes dominate the population.

Finally, Fig.~\ref{inclinazione} reports the distribution of the taxa with respect to
the orbital inclination $i$.
A 5$\degr$ binning is adopted (two objects are out of the scale).
Inclinations of RR-types are quite low, in agreement with the previously mentioned finding of a red dynamically ``cold'' population.  
On the contrary, BB-types seem to be concentrated at high inclinations, confirming the suggested association
of these objects with the ``hot'' population (Levison \& Stern 2001; Brown 2001; Doressoundiram et al. 2002).

\begin{table*}[p]
\caption{Average colors of the four taxa.}
\label{colors}      % is used to refer this table in the text
{\centering
\begin{tabular}{l c c c c c c}        % centered columns (4 columns)
\hline\hline                 % inserts double horizontal lines
Class                       & $B-V$		& $V-R$		& $V-I$		& $V-J$		& $V-H$		& $V-K$          \\
\hline                        % inserts single horizontal line
BB                          & $0.68\pm0.06$     & $0.39\pm0.05$ & $0.75\pm0.06$ & $1.20\pm0.25$ & $1.28\pm0.50$ & $1.32\pm0.60$  \\
BR                          & $0.75\pm0.06$     & $0.49\pm0.05$ & $0.96\pm0.08$ & $1.68\pm0.15$ & $2.11\pm0.14$ & $2.26\pm0.13$  \\
IR                          & $0.92\pm0.05$     & $0.59\pm0.04$ & $1.15\pm0.07$ & $1.86\pm0.06$ & $2.20\pm0.09$ & $2.29\pm0.10$  \\
RR                          & $1.06\pm0.10$     & $0.69\pm0.06$ & $1.35\pm0.10$ & $2.22\pm0.20$ & $2.58\pm0.25$ & $2.59\pm0.28$  \\
\hline                                   %inserts single line
\end{tabular}
~\\
}
%\raggedright
\smallskip
NOTE: All of the 151 Centaurs and TNOs that have been classified thus far are taken into account.
In the cases where multiple taxonomic classes were assigned to an object, we considered the first designation.
\end{table*}

\section{Conclusions}
During the second year of an ESO Large Programme on TNOs and Centaurs,
photometric observations of 31 objects have been carried out.
From the comparison with previous works, hints of heterogeneous surfaces have been found for 4 objects
(Chariklo, Sedna, 2005 RM$_{43}$, and 2005 RR$_{43}$).

For 28 out of the 31 observed objects we derived the taxomomic
classification (within the system by Barucci et al. 2005), using G-mode analysis.
Taxonomy for ten additional objects was obtained in the framework of our programme (DeMeo et al. 2009a),
for a total of 38 TNOs and Centaurs. This sample includes 19 objects which are classified for the first time ever,
which constitutes about a 14\% increase of the sample of 132 objects analysed by Fulchignoni et al. (2008)
using the available literature.

We took into account the 151 objects that have been classified thus far
to compute the average colors of the four taxonomic groups
and to analyse their distribution with respect to the dynamical properties of the objects.

Looking at the distribution of the taxa within dynamical classes, the already known color bimodality (BR, RR)
of Centaurs clearly emerges. Also, all of the four objects we classified as IR belong to the classical TNOs, in
agreement with the finding that IR-types are concentrated in the resonant and classical dynamical classes
(Fulchignoni et al. 2008).
RR-types dominate the classical TNOs, but a similar division in the four taxa appears among the classical TNOs
classified in the framework of our programme. 

BR and RR classes dominate among the population at small ($a \la 30\degr$) values of semimajor axis, while all the four taxa
are well represented at greater distances from the Sun.

RR and BB classes are more abundant at low and high orbital inclinations, respectively,
which associates these objects with the dynamically ``cold'' and ``hot'' populations.

%\begin{acknowledgements}
%\end{acknowledgements}

\end{document}